\documentstyle[12pt,epsfig]{article}
\begin{document}

\title{
\begin{flushright} \large
ATLAS Internal Note \\
PHYS-NO-72 \\
10 October 1995
\end{flushright}
\vspace{1.5cm}
\Large
$B_s^0 \to D_s^-a_1^+$ decay channel in the $B_s^0$-mixing studies}
\protect\author{A.V.~Bannikov, G.A.~Chelkov, Z.K.~Silagadze\thanks
{ permanent address: Budker Institute of Nuclear Physics,
630 090, Novosibirsk, Russia.}
\vspace*{3mm} \\
\sc JINR, Dubna \vspace*{3mm} \\}
\date{}
\maketitle

\begin{abstract}
The usefulness of the $B_s^0 \to D_s^-a_1^+, D_s^- \to \phi \pi^- \to
K^+K^-\pi^-, a_1^+ \to \rho^0 \pi^+ \to \pi^+ \pi^- \pi^+$ decay chain
is investigated for the $B_s$-reconstruction in the future ATLAS
$B_s$-mixing experiment. It is shown that this decay channel is almost
as suitable for this purpose as previously studied $B_s^0 \to D_s^-\pi^+$.
\end{abstract}

\section{Introduction}

In the Standard Model $B^0$ and $\bar {B^0}$ are not mass eigenstates.
Instead we have (the small CP-violating effects are neglected)
\begin{eqnarray}
B^0=\frac{B_1+B_2}{\sqrt{2}} \hspace*{1cm}
\bar {B^0}=\frac{B_1-B_2}{\sqrt{2}}
\; . \label{eq1} \end{eqnarray}
So the time evolution of the $B_i$ states
looks like
\begin{eqnarray}
B_i(t)=B_i(0) exp \left \{-\frac{i}{\hbar}(m_i-i\frac{\Gamma_i}{2})t
\right \} \; \; , \label{eq2} \end{eqnarray}
\noindent
where $m_i$ is the mass eigenvalue and $\Gamma_i$ - the corresponding width.

It follows from (1) and (2) that the probability for $B^0$ meson not
to change its flavour after a time $t$ from the creation is
\begin{eqnarray}
P^{B^0B^0}(t)=|\langle B^0(t)|B^0(0)\rangle |^2 =
\frac{1}{2} e^{-\frac{\Gamma t}{\hbar}}\left ( \cosh{\frac{\Delta
\Gamma t}{2\hbar}}+\cos{\frac{\Delta m \, t}
{\hbar}} \right ) \; \; ,
\label{eq3} \end{eqnarray}
\noindent
and the probability to convert into the $\bar {B^0}$ meson --
\begin{eqnarray}
P^{B^0\bar {B^0}}(t)=|\langle \bar {B^0}(t)|B^0(0)\rangle |^2 =
\frac{1}{2} e^{-\frac{\Gamma t}{\hbar}}\left ( \cosh{\frac{\Delta
\Gamma t}{2\hbar}}- \cos{\frac{\Delta m \, t}
{\hbar}} \right ) \; \; ,
\label{eq4} \end{eqnarray}\noindent
where $\Gamma=\frac{1}{2}(\Gamma_1+\Gamma_2)$ is the average width and
$\Delta \Gamma=\Gamma_1-\Gamma_2 \, $.
So $\Delta m=m_1-m_2$ mass difference between the $B$ mass eigenstates
defines the oscillation frequency. Standard Model predicts \cite{1}
that $\frac{\Delta m_s}{\Delta m_d} \sim \left \vert \frac{V_{ts}}{V_{td}}
\right \vert ^2 \gg 1$, $V_{ij}$ being the
Cabibbo-Kobayashi-Maskawa matrix element.
Therefore the mixing in the $B_s^0$ meson system proceeds much more
faster than in the $B_d^0$ system.

The total probability $\chi$ that a $B^0$ will oscillate into $\bar {B^0}$ is
\begin{eqnarray}
\chi = \int\limits_0^\infty P^{B^0 \bar {B^0}} (t) \, \frac{dt}{\tau}
(1-y^2) = \frac{1}{2} \left ( \frac{x^2}{1+x^2} +
\frac{y^2}{1-y^2} \right ) (1-y^2) \hspace*{3mm} , \;
x=\frac{\Delta m}{\Gamma} \; , \; y=\frac{\Delta \Gamma}{2\Gamma}
\; . \label{eq5} \end{eqnarray}
In the first $B_d$-mixing experiments \cite{2} just this time integrated
mixing probability was measured. The result  \cite{3} $x_d=0.69 \pm 0.07$
shows that in the $B_s$ system $x_s \gg 1$ is expected. In fact the
allowed range of $x_s$ is estimated to be between $\sim 12$ and $\sim 30$
in the Standard Model \cite{4}. Such a big value of $x_s$ makes impossible
time integrated measurements in the $B_s$ system, because $\chi$ in (5)
saturates at $\sim 0.5$ for large values of x.

Although it was thought that unlike the kaon system for the $B$ mesons
the decay width difference can be neglected \cite{5}, nowadays people
is more inclined to believe the theoretical prediction \cite{6} that
the $b \to c \bar c s$ transition, with final states common to both
$B_s$ and $\bar B_s$, can generate about 20\% difference in lifetimes
of the short lived and long lived $B_s$-mesons \cite{7}.

But we can see from the (3 $\div$ 5) formulas that the effect of nonzero
$y$ is always $\sim y^2$ and so of the order of several percents, because
$y \approx 0.1$ is expected. In the following we will neglect this effect
and will take $y=0$, though in some formulas $y$ is kept for reference
reason.

The development of high precision vertex detectors made it possible to
measure \cite{8} in the $B_d$ system the time dependent asymmetry
\begin{eqnarray}
\frac{P^{B^0B^0}-P^{B^0 \bar{B^0}}}{P^{B^0B^0}+P^{B^0 \bar{B^0}}} =
\cos {\frac{\Delta m \, t}{\hbar}} \;\; .
\label{eq6} \end{eqnarray}
The same techniques can also be applied to the $B_s-\bar B_s$ system.

Recently the ATLAS detector sensitivity to the $x_s$ parameter was studied
\cite{9} using $B_s^0 \to D_s^-\pi^+ \to \phi \pi^- \pi^+ \to
K^+K^-\pi^-\pi^+$ decay chain for $B_s$ meson reconstruction. It was shown
that $x_s$ up to 40 should be within a reach \cite{10}.
The signal statistics could be increased by using other decay channels,
like $B_s^0 \to D_s^-a_1^+ \; , \; B_s^0 \to J/\Psi K^{*0}$.

The purpose of this note is to study the usefulness of the decay chain
$B_s^0 \to D_s^-a_1^+ \to \phi \pi^- \rho \pi^+ \to K^+K^-\pi^-\pi^+
\pi^-\pi^+$ for $B_s$ meson reconstruction in the ATLAS $B_s$-mixing
experiments.

\section{Event simulation}

About 20 000 following b-decays were generated using the PYTHIA Monte
Carlo program \cite{11}

\begin{center}
\begin{tabbing}
$ (p^{\mu}_T > 6~GeV/c, \mid \eta^{\mu} \mid < 2.2)~
 \mu_{tag} \leftarrow b\bar{b} \rightarrow B^0_s \rightarrow$
\= $D^-_s$ \= $a^+_1$  \\
\> \> $\hookrightarrow$ \= $\rho^0 \pi^+$ \\
\> \> \> $\hookrightarrow \pi^+ \pi^-$ \\
\> $\hookrightarrow$ \= $\phi \pi^-$ \\
\> \> $\hookrightarrow K^+K^-$
\end{tabbing}
\end{center}

The impact
parameter was smeared using the following parameterized description of
the impact parameter resolution
\begin{eqnarray}
\sigma_{IP}=14 \oplus 72/(p_T \sqrt{|\sin {\theta}|}) \hspace*{5mm}
\sigma_{Z}=20 \oplus 83/(p_T \sqrt{|\sin {\theta}|^3})
\hspace*{3mm} ,
\label{eq7} \end{eqnarray}
\noindent
where resolutions are in $\mu m$ and $\theta$ is the angle with respect
to the beam line. It was shown in \cite{9} that this parameterized
resolution reasonably reproduces the results obtained by using the full
simulation and reconstruction programs.

For the transverse momentum resolution an usual expression \cite{10}
\begin{eqnarray}
\frac{\sigma (p_T)}{p_T}=5 \cdot 10^{-4} p_T \oplus 1.2 \%
\label{eq8} \end{eqnarray} \noindent
was assumed.

Track reconstruction efficiencies for various particles were taken
from \cite{10}. Because now we have 6 particles in the final state instead
of 4 for the $B_s^0 \to D_s^-\pi^+$ decay channel, we expect some loss
in statistics due to track reconstruction inefficiencies, but the effect
is not significant because the investigation in \cite{10} indicates a
high reconstruction efficiency of 0.95.

\section{Event reconstruction}
 The topology of a considered $B_s^0$ decay chain is shown schematically
in a figure:

\setlength{\unitlength}{1mm}\thicklines
\begin{picture}(150,110)
\put(40,40){\circle*{4}}
\put(40,40){\vector(-3,-2){30}}
\put(12,17){{\Large\bf $\mu_{tag}$}}
\put(40,40){\line(2,1){40}}
\put(62,45){{\Large\bf $B^0_s$}}
\put(80,60){\line(3,-1){15}}
\put(83,52){{\Large\bf $D^-_s$}}
\put(95,55){\vector(3,1){33}}
\put(130,65){{\Large\bf $K^+$}}
\put(95,55){\vector(3,0){35}}
\put(133,55){{\Large\bf $K^-$}}
\put(95,55){\vector(3,-2){35}}
\put(130,35){{\Large\bf $\pi^-$}}
\put(80,60){\vector(1,3){10}}
\put(85,90){{\Large\bf $\pi^+$}}
\put(80,60){\vector(2,3){20}}
\put(102,90){{\Large\bf $\pi^-$}}
\put(80,60){\vector(3,2){30}}
\put(112,80){{\Large\bf $\pi^+$}}
\end{picture}

The $B_s$ decay vertex reconstruction was done in the following
three steps.

First of all the $D_s^-$ was reconstructed by finding three charged
particles presumably originated from the $D_s^-$ decay and fitting their
tracks. For this goal all combinations of the properly charged particles
were examined in the generated events, assuming that two of them are kaons
and one is pion. The resulting invariant mass distribution is shown in
Fig.1a for signal events. The expected $D_s^-$ peak is clearly seen along
with moderate
enough combinatorial background. Cuts on $\Delta \phi_{KK}$~,~ $\Delta
\Theta_{KK}$ and $|M_{KK}-M_\phi|$ were selected in order to optimize
signal to background ratio. To select one more cut on $|M_{KK\pi}-M_
{D_s^-}|$, the information about the invariant mass resolution is desirable.
Fig. 2a shows the reconstructed $D_s^-$ meson from its true decay products.
The finite invariant mass resolution is due to applied track smearing and
equals approximately to $10 MeV/c^2$.

After $D_s^-$ meson reconstruction, $a_1^+$ meson was searched in three
particle combinations from the remaining charged particles, each particle
in the combination being assumed to be a pion. Fig. 1b shows a resulting
invariant mass distribution for signal events. Because of huge width of
$a_1^+$, signal to background
separation is not so obvious in this case. If $a_1^+$ is reconstructed from
its true decay products as in Fig. 2b, its width is correctly reproduced.
To draw out $a_1^+$ from the background, further cuts were applied on
$\Delta \phi_{\pi \pi}$~,~ $\Delta \Theta_{\pi \pi}$~,~
$|M_{\pi \pi}-M_\rho|$  and $|M_{\pi \pi \pi}-M_{a_1}|$.

At last $B_s^0$ decay vertex was fitted, using reconstructed $D_s^-$
and $a_1^+$.

Almost the same resolution in the $B_s$-decay proper time was reached
$\sigma_\tau \approx 0.064 ps$, as in \cite{9}. The corresponding
resolution in the B-meson decay length in the transverse plane is
$\approx 87 \mu m$. The relevant distributions are shown
in Fig.3.

\section{Signal and background}

Branching ratios and signal statistics for the $B_s^0 \to D_s^- a_1^+$
channel are summarized in Table 1. Note that we use an updated value for
Br($D_s^- \to \phi \pi^-$) from \cite{12}. $B_s^0$ branching ratios are
still unknown experimentally. Neglecting SU(3) unitary symmetry breaking
effects, we have taken Br($B_s^0 \to D_s^- a_1^+$)$\approx$
Br($B^0 \to D^- a_1^+$).


\begin{center}
Table 1.\\
Branching ratios and signal statistics
for $B_s^0\rightarrow D_s^{-}a_1^{+}(1260)$.\\
\vspace{0.5cm}
\begin{tabular}{|l|c|l|}
\hline
\hline
Parameter & Value & Comment \\
\hline
\hline
  $L~[cm^{-2}s^{-1}]$                          & $10^{33}$  &   \\
  $t~[s]$                                      & $10^7$     &   \\
  $\sigma(b \bar b)/\sigma(tot)$               & $\simeq 1/100$ &   \\
  $\sigma(b \bar b)~[\mu b]$                   & $\simeq 500$   &   \\
  $\sigma(b \bar b \rightarrow \mu X)~[\mu b]$ & $\simeq 2.24$   &
  $p^{\mu}_{T} > 6~GeV/c$  \\
					       &            &
  $|\eta^{\mu}| < 2.2$     \\
\hline
  $N(b \bar b \rightarrow \mu X)$              & $2.24 \times 10^{10}$ &  \\
\hline
  $Br(b \rightarrow B^0_s)$                    & $0.1$      &   \\
  $Br(B^0_s \rightarrow D^-_{s}a_1^+)$         & $0.006$    &   \\
  $Br(D^-_s \rightarrow \phi\pi^-)$            & $0.035$    &   \\
  $Br(\phi \rightarrow K^+K^-)$                & $0.491$    &   \\
  $Br(a_1^+ \rightarrow \rho^0\pi^+)$          & $\sim 0.5$ &   \\
  $Br(\rho^0 \rightarrow \pi^-\pi^+)$          & $\sim 1$   &   \\
\hline
  $N(K^+K^-\pi^-\pi^+\pi^-\pi^+)$              & $116000$    &   \\
\hline
\hline
\end{tabular}
\end{center}
\vspace{0.5cm}


Acceptance and analysis cuts are summarized in Table 2. We take
a track reconstruction efficiency of 95\% and a lepton identification
efficiency of 80\%, as in \cite{10}.

\pagebreak

\begin{center}
Table 2.\\
Analysis cuts and acceptance for $B_s^0\rightarrow D_s^{-}a_1^{+}(1260)$
(for $10^4 \, pb^{-1}$ integrated luminosity).\\
\vspace{0.5cm}
\begin{tabular}{|l|c|l|}
\hline
\hline
Parameter & Value & Comment \\
\hline
\hline
  $N(K^+K^-\pi^-\pi^+\pi^-\pi^+)$              & $116000$    &   \\
\hline
  Cuts :                         &            &   \\
  $p_{T} > 1~GeV/c$                            &            &   \\
  $|\eta| < 2.5$                               &            &   \\
\hline
  $N(K^+K^-\pi^-\pi^+\pi^-\pi^+)$              & $7680$     & $6.6\%$  \\
\hline
  $\Delta\varphi_{KK} < 10^\circ$              &            &   \\
  $\Delta\theta_{KK} < 10^\circ$               &            &   \\
  $|M_{KK}-M_{\phi}| < 20~MeV/c^2$             &            &   \\
  $|M_{KK\pi}-M_{D_s^-}| < 15~MeV/c^2$         &            &   \\
  $\Delta\varphi_{\pi\pi} < 35^\circ$          &            &   \\
  $\Delta\theta_{\pi\pi} < 15^\circ$           &            &   \\
  $|M_{\pi\pi}-M_{\rho^0}| < 192~MeV/c^2~(\pm 3\sigma) $    &     &   \\
  $|M_{\pi\pi\pi}-M_{a_1^+}| < 300~MeV/c^2$    &            &   \\
\hline
  $N(K^+K^-\pi^-\pi^+\pi^-\pi^+)$              & $5765$     & $5.0\%$  \\
\hline
  $D_s^{-}$ vertex fit $\chi^2 < 12.0$         &            &   \\
  $a_1^{+}$ vertex fit $\chi^2 < 12.0$         &            &   \\
  $B_s^{0}$ vertex fit $\chi^2 < 0.35$         &            &   \\
  $B_s^{0}$ proper decay time $> 0.4~ps$       &            &   \\
  $B_s^{0}$ impact parameter $< 55~\mu m$      &            &   \\
  $B^0_s~~p_T > 10.0~GeV/c$                    &            &   \\
\hline
  $N(K^+K^-\pi^-\pi^+\pi^-\pi^+)$ after cuts   & $3505$     & $3.0\%$  \\
\hline
  Lepton identification                        & $0.8$      &   \\
  Track efficiency                             & $(0.95)^6$ &   \\
\hline
\hline
  $N(K^+K^-\pi^-\pi^+\pi^-\pi^+)$ reconstructed  & $2065$   & $1.8\%$  \\
\hline
\hline
\end{tabular}
\end{center}


As we see, about 2065 reconstructed $B_s^0$ are expected after one year
run at ${\cal L}=10^{33} cm^{-2} s^{-1}$ luminosity. The corresponding
number of events within one standard deviation \- ($\simeq 22~ MeV/c^2$)
from the $B_s^0$ mass equals 1407. This last number should be compared
to 2650 signal events, as reported in \cite{10},
when $B_s^0 \to D_s^- \pi^+$ decay channel is used.

Events which pass the first level muon trigger ($p_T > 6~ GeV/c,~
|\eta|<2.2$) are predominantly $b \bar b$ events. Background can come
from other $B$ decays of the same or higher charged multiplicity, and from
random combinations with some (or all) particles originating not from
a $B$ decay (combinatorial background).

The following channels were considered and no significant contributions
were found to the background:
\begin{itemize}
\item $B^0_d \to D^- a_1^+$. These events don't pass the analysis cuts, because
the $D^-$ mass is shifted from the $D_s^-$ mass by about 100 MeV, and so
does the $B_d^0$ mass compared to the $B_s^0$ mass.
\item $\Lambda_b \to \Lambda_c^+ \pi^-$ followed by $\Lambda_c^+ \to
pK^- \pi^+ \pi^+ \pi^-$ . Taking $Br(\Lambda_b \to \Lambda_c^+ \pi^-)
\approx 0.01$ from \cite{13}, we see that the expected number of
$pK4\pi$ events, originated from this source, is only five times less
than the expected number of truly signal events. But the decay topology
for this decay chain is drastically different (1+5, not 3+3) and therefore
it is unexpected that significant amount
of the B-decays will be simulated in this way.

Note that even for $B_s \to D_s^-\pi^+$ decay channel the similar background
is negligible \cite{9}, although $Br(\Lambda_c^+ \to pK^- \pi^+ )$ is
about 44 times bigger than $Br(\Lambda_c^+ \to pK^- \pi^+ \pi^+ \pi^-)$.
\item $B^0_d \to D_s^- a_1^+$. About 10 000 such events were generated by
PYTHIA and then analyzed. Using Br($B_d^0 \to D_s^+ a_1^-)<2.7 \cdot
10^{-3} $ from \cite{12} and assuming that $B_d^0 \to D_s^- a_1^+$ decay
goes through $B^0 \bar {B^0}$ oscillations: $B_d^0 \to \bar{B_d^0} \to
D_s^- a_1^+$, and therefore
$Br(B_d^0 \to D_s^- a_1^+)=\chi_d Br(B_d^0 \to D_s^+ a_1^-)<4.3 \cdot
10^{-4}$, we have got Fig.4. It is seen from this figure that because of
$M_{B_s}-M_{B_d} \approx 100 MeV$ mass shift, the contribution of this
channel to the background proves to be negligible.

Note that Fig.4 refers to the total number of the $B^0_d \to D_s^- a_1^+$
events. In fact the distribution of these events with regard to the decay
proper time is oscillatory, $x_d$ (not $x_s$) defining the oscillation
frequency. So in general this will result in oscillatory dilution factor.
The conclusion that this dilution factor is irrelevent relies on the fact
that no candidate event was found with invariant mass within one standard
deviation from the $B_s$ mass for $6\cdot 10^4 \, pb^{-1}$ integrated
luminosity.
\end{itemize}

A huge Monte-Carlo statistics is needed for combinatorial background
studies. No candidate event with $M_{B_s^0}-150MeV/c^2 <
M_{KK\pi \pi \pi \pi} < M_{B_s^0}+150MeV/c^2$
was found within $\sim 3\cdot 10^5$ inclusive $\mu X$ events. This indicates
that signal/background ratio is expected to be not worse than 1:1.

\section{Dilution factors}
The observation of the $B-\bar B$ oscillations is
complicated by some dilution factors. First of all the decay proper time
is measured with some accuracy $\sigma$. From previous discussions we
know that in our case $\sigma = 0.064 ps$ is expected. Due to this
finite time resolution, the observed oscillations are convolutions of the
expressions (3) and (4) given above with a Gaussian distribution. For
example
\begin{eqnarray} &&
P^{B^0 \bar {B^0}} \to \frac{1}{2} \int \limits_{-\infty}^\infty
e^{-\frac{\Gamma s}{\hbar}} \left ( \cosh{\frac{\Delta
\Gamma t}{2\hbar}}-\cos{\frac{\Delta m s}{\hbar}}
\right )\exp{\left [ -\frac{(t-s)^2}{2\sigma^2} \right ] }
\frac{ds}{\sqrt{2\pi}\sigma} \sim \nonumber \\ &&
\frac{1}{2} e^{-\frac{\Gamma t}{\hbar}}\left ( \cosh{\frac{\Delta
\Gamma }{2\hbar}}(t-\sigma \frac{\sigma}{\tau})-D_{time}\cos
{\frac{\Delta m}{\hbar}(t-\sigma \frac{\sigma}{\tau})} \right ) \; \; ,
\label{eq9} \end{eqnarray}
\noindent
where $D_{time}=\exp{\left [-\frac{1}{2} (\frac{\sigma}{\tau} )^2
(x^2+y^2)\right ] },~\tau = \frac{\hbar}{\Gamma}$.

So the main effect of this smearing is the reduction of the oscillation
amplitude by $D_{time}$. This is quite important in the $B_s$ system where
$x \gg 1$. There is also a time shift $t \to t-\sigma
\frac{\sigma}{\tau}$ in (9). This time shift does not really effect the
observability of the oscillations and we will neglect it.

In fact (9) is valid only for not too short decay times $t \gg \sigma$,
because in (3) and (4) distributions $t>0$ is assumed.

Another reduction in the oscillation amplitude is caused by the particle/
antiparticle mistagging at t=0. In our case particle/antiparticle nature
of the $B$ meson is tagged by the lepton charge in the semileptonic decay
of the associated beauty hadron. Mistagging is mainly due to
\begin{itemize}
\item $B- \bar B$ oscillations: accompanying b-quark can be hadronized as a
neutral $B$ meson and oscillate into $\bar B$ before semileptonic decay.
\item $b \to c \to l^+$ cascade process, then the lepton is misidentified
as having come directly from the $B$-meson and associated to the
$\bar b \to l^+$ decay.
\item leptons coming from other decaying particles (K,$\pi$,...).
\item detector error in the lepton charge identification.
\end{itemize}

Let $\eta$ be the mistagging probability. If we have tagged $N$ $B^0$
mesons, among them only $(1-\eta)N$ are indeed $B^0$-s and $\eta N$
are $\bar {B^0}$-s misidentified as $B^0$-s. So at the proper time $t$
we would observe ($P^{\bar {B^0} \bar {B^0}}(t)=P^{B^0 B^0}(t)$ due to
CPT invariance)
\begin{eqnarray}
N \left [ (1-\eta)P^{B^0 \bar {B^0}}(t)+\eta P^{B^0 B^0}(t) \right ]=
\frac{N}{2}e^{-\frac{\Gamma t}{\hbar}} \left [\cosh{\frac{\Delta
\Gamma t}{2\hbar}}-(1-2\eta)
\cos {\frac{\Delta m \, t}{\hbar}} \right ]
\nonumber \end{eqnarray}
\noindent
decays associated to the $\bar {B^0}$ meson and therefore
$$P^{B^0 \bar {B^0}}(t) \to
\frac{1}{2}e^{-\frac{\Gamma t}{\hbar}} \left [\cosh{\frac{\Delta
\Gamma t}{2\hbar}}-(1-2\eta)
\cos {\frac{\Delta m \, t}{\hbar}} \right ] \; \; \; . $$
So the dilution factor due to mistagging is $D_{tag}=1-2\eta$.
In our studies we have taken $D_{tag}=0.56$, as in \cite{14}.

Finally the dilution can emerge from background. Suppose that apart
from  $$\frac{N_{signal}}{2} e^{-\frac{\Gamma t}{\hbar}} \left
(\cosh{\frac{\Delta
\Gamma t}{2\hbar}}-
\cos {\frac{\Delta m \, t}{\hbar}} \right )$$ events with $B \to \bar B$
oscillations we also have $N_{back}(t)$ additional background events.
Half of them will simulate $\bar B$ meson and half of them B meson
(assuming asymmetry free background). So
the observed number of would be $B \to \bar B$ oscillations will be
$$\frac{N_{signal}}{2}e^{-\frac{\Gamma t}{\hbar}} \left (\cosh{\frac{\Delta
\Gamma t}{2\hbar}}-
\cos {\frac{\Delta m \, t}{\hbar}} \right ) +\frac{N_{back}(t)}{2}
\sim e^{-\frac{\Gamma t}{\hbar}} \left (\cosh{\frac{\Delta
\Gamma t}{2\hbar}}-
D_{back} \cos {\frac{\Delta m \, t}{\hbar}} \right )
$$
\noindent
and the oscillation amplitude will be reduced by an amount
$$ D_{back}=\frac{N_{signal} \cdot \cosh{\frac{y
t}{\tau}}}{N_{signal}+N_{back}(t)e^{\frac{t}
{\tau}}} \; \; \; .$$
Neglecting the proper time dependence of this dilution factor
(that is supposing that the background is mainly due to $B$-hadron
decays and therefore has approximately the same proper time exponential
decay as the signal \cite{15}),we have
taken $D_{back} \approx 0.71$ which corresponds to the 2:1
signal/background ratio.

\section{Prospects for $x_s$ measurements}
For $6 \cdot 10^4 \; pb^{-1}$ integrated luminosity
 the number of reconstructed
$B_s^0$-s would reach $\sim 8000$ from the analyzed channel alone.
Another $\sim 16 000 \; \; B_s^0$-s are expected from the $B_s^0 \to
D_s^- \pi^+$ channel \cite{9,10}.

For events in which $B_s^0$ meson does not oscillate before its decay,
the $D_s$ meson and the tagging muon have equal sign charges. If the
$B_s^0$ meson oscillates, opposite charge combination emerges. The
corresponding decay time distributions are
\begin{eqnarray} &&
\frac{dn(++)}{dt}=\frac{N}{2\tau} e^{-\frac{t}{\tau}}\left ( 1+D\cos
{(\frac{x_s t}{\tau})}  \right ) \nonumber \\ &&
\frac{dn(+-)}{dt}=\frac{N}{2\tau} e^{-\frac{t}{\tau}}\left ( 1-D\cos{
(\frac{x_s t}{\tau})}  \right )
\label{eq10} \end{eqnarray}
\noindent
D is the product of all dilution factors and $N$ is the total number of
reconstructed $B_s^0$-s.

The unification of samples from $B_s^0 \to D_s^- a_1^+$ and
$B_s^0 \to D_s^- \pi^+$ decay channels allows to increase $x_s$
measurement precision.

Fig.7 and Fig.8 show the corresponding
$$ A(t)=\frac{\frac{dn(++)}{dt}-\frac{dn(+-)}{dt}}
{\frac{dn(++)}{dt}+\frac{dn(+-)}{dt}}=D\cos{(\frac{x_s t}{\tau})} $$
asymmetry plots for $x_s=20$ and 35.

\section{Conclusions}
It seems to us that $B^0_s \to D_s^-a_1^+$ decay channel is almost as
good for the $B_s$-mixing exploration as previously studied
$B^0_s \to D_s^-\pi^+$ and enables us to increase signal statistics
about 1.5 times. Further gain in signal statistics can be reached \cite{9,10}
by using $B_s^0 \to J/\psi K^*$ decay mode and considering other decay
channels of $D_s^-$. These possibilities are under study.

We refrain from giving any particular value of $x_s$ as an attainable
upper limit. Too many uncertainties are left before a real experiment
will start. Note, for example, that about two times bigger branching
ratios for both $B_s \rightarrow D^-_s \pi^+$ and
$B_s \rightarrow D^-_s a^+_1$ decay channels are predicted in \cite{16}.
$\sim 500 \mu b$ as a $b \bar b$ production cross section can also
have significant variation in real life \cite{17}.

So although the results of this investigation strengthen confidence
in reaching $x_s$ as high as 40 \cite{10}, it should be realized that
some theoretical predictions about $B_s$-physics and collider
operation were involved and according to T.D.Lee's first law of
physicist \cite{18} "without experimentalist, theorist tend to
drift". However maybe it is worthwhile to recall his second law
also "without theorist, experimentalists tend to falter".

\section*{Acknowledgements}
Many suggestions of P.Eerola strongly influenced this investigation
and lead to considerable improvement of the paper. Communications
with S.Gadomski and N.Ellis are also appreciated.
Authors thank N.V.~Makhaldiani for drawing their attention to
T.D.Lee's paper.

\begin{figure}
\vspace*{2cm}
\epsfig
{file=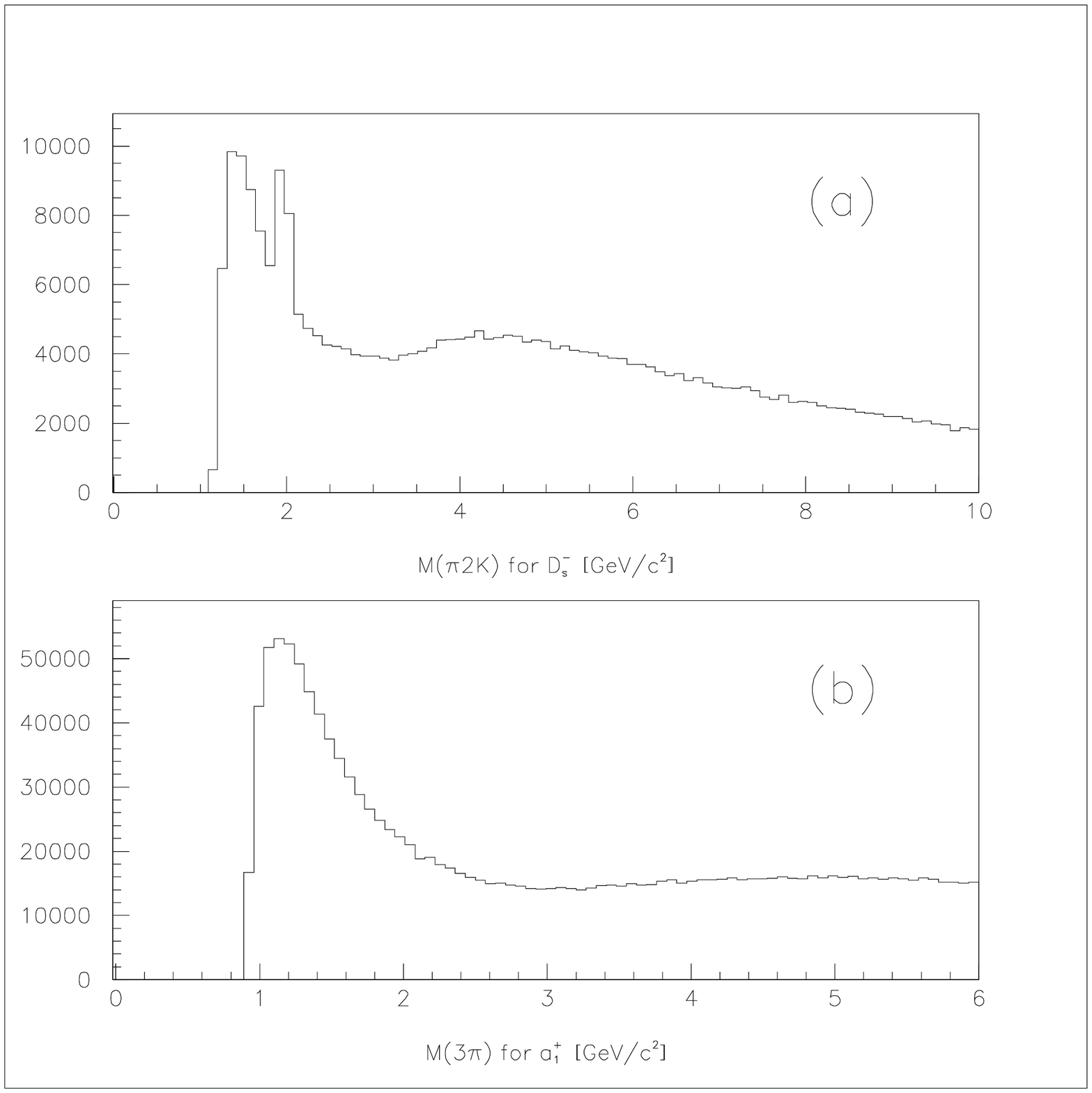,height=13cm,bbllx=40pt,bblly=70pt,bburx=625pt,bbury=550pt}
\caption{Invariant mass distributions of three charged particle combinations
in signal events,
assuming $2K+\pi$ (a) or $3\pi$ combination (b) as described in the text.}
\end{figure}

\begin{figure}
\epsfig
{file=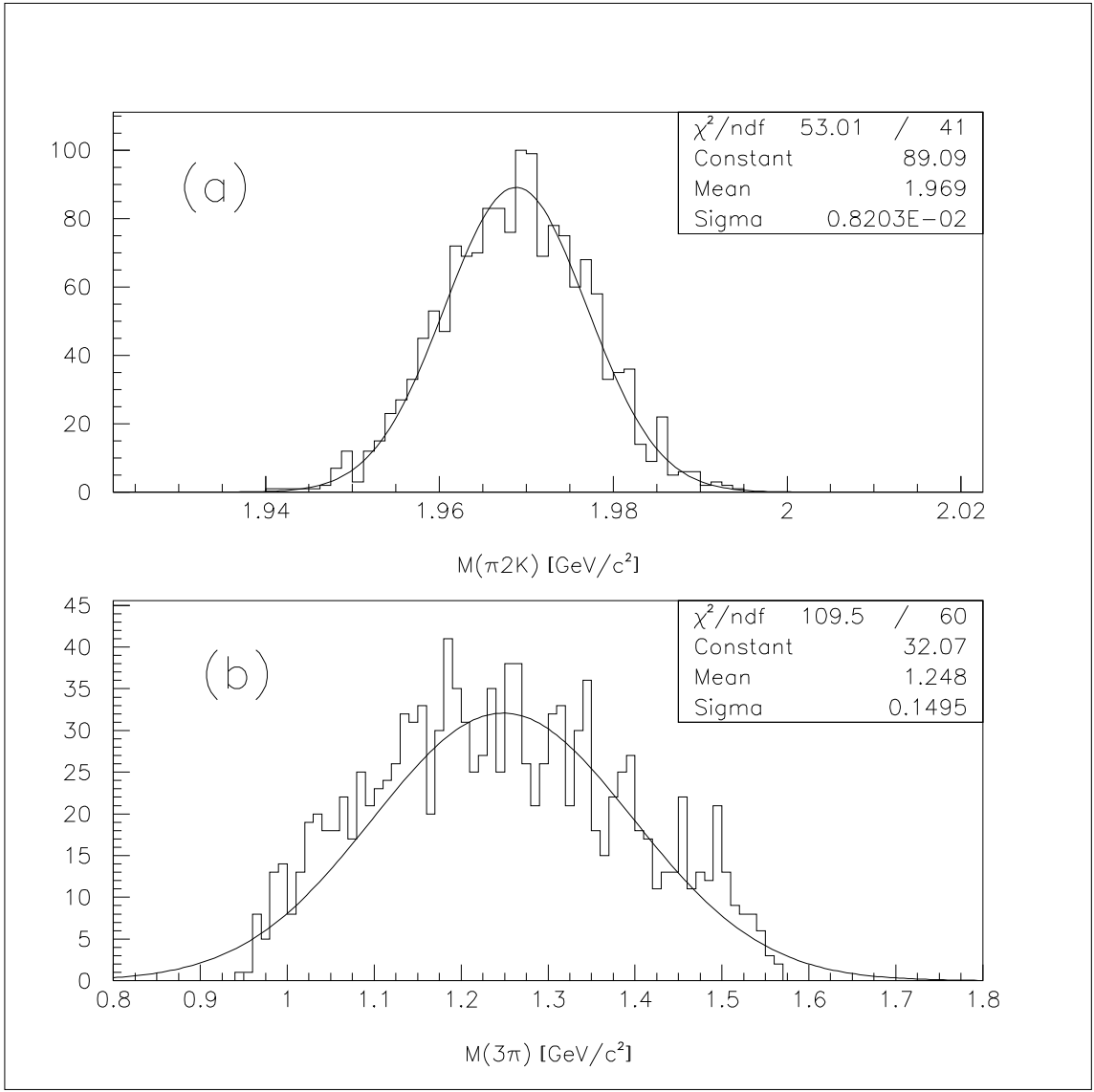,height=13cm,bbllx=40pt,bblly=70pt,bburx=625pt,bbury=550pt}
\caption{Three particle invariant mass distributions
of reconstructed $D^-_s$ and $a_1^+$ events.}
\end{figure}

\begin{figure}
\epsfig
{file=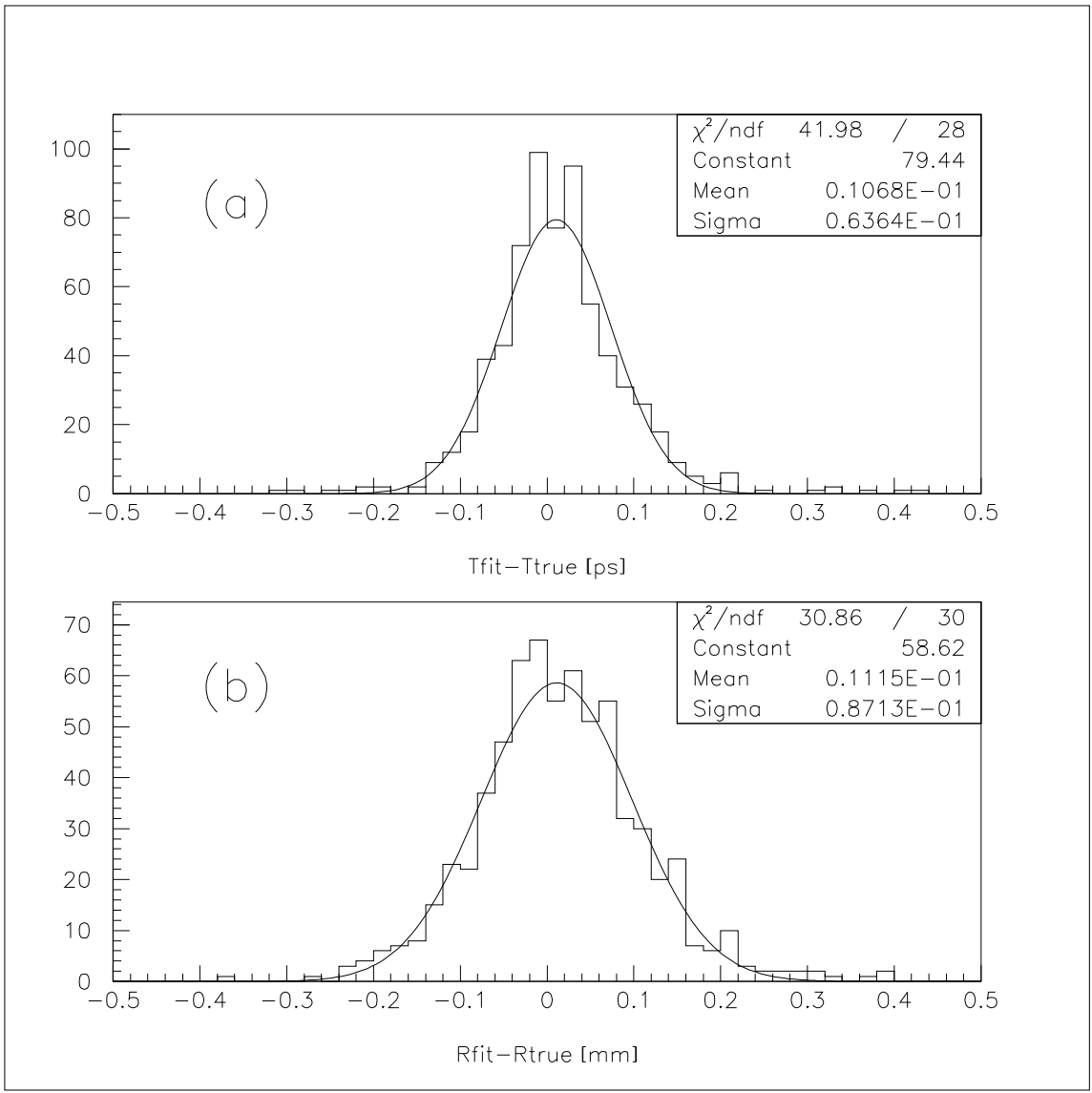,height=13cm,bbllx=40pt,bblly=70pt,bburx=625pt,bbury=550pt}
\caption{Proper time (a) and transverse radius (b) resolutions for the
reconstructed $B_s^0$ decay vertex.}
\end{figure}

\begin{figure}
\epsfig
{file=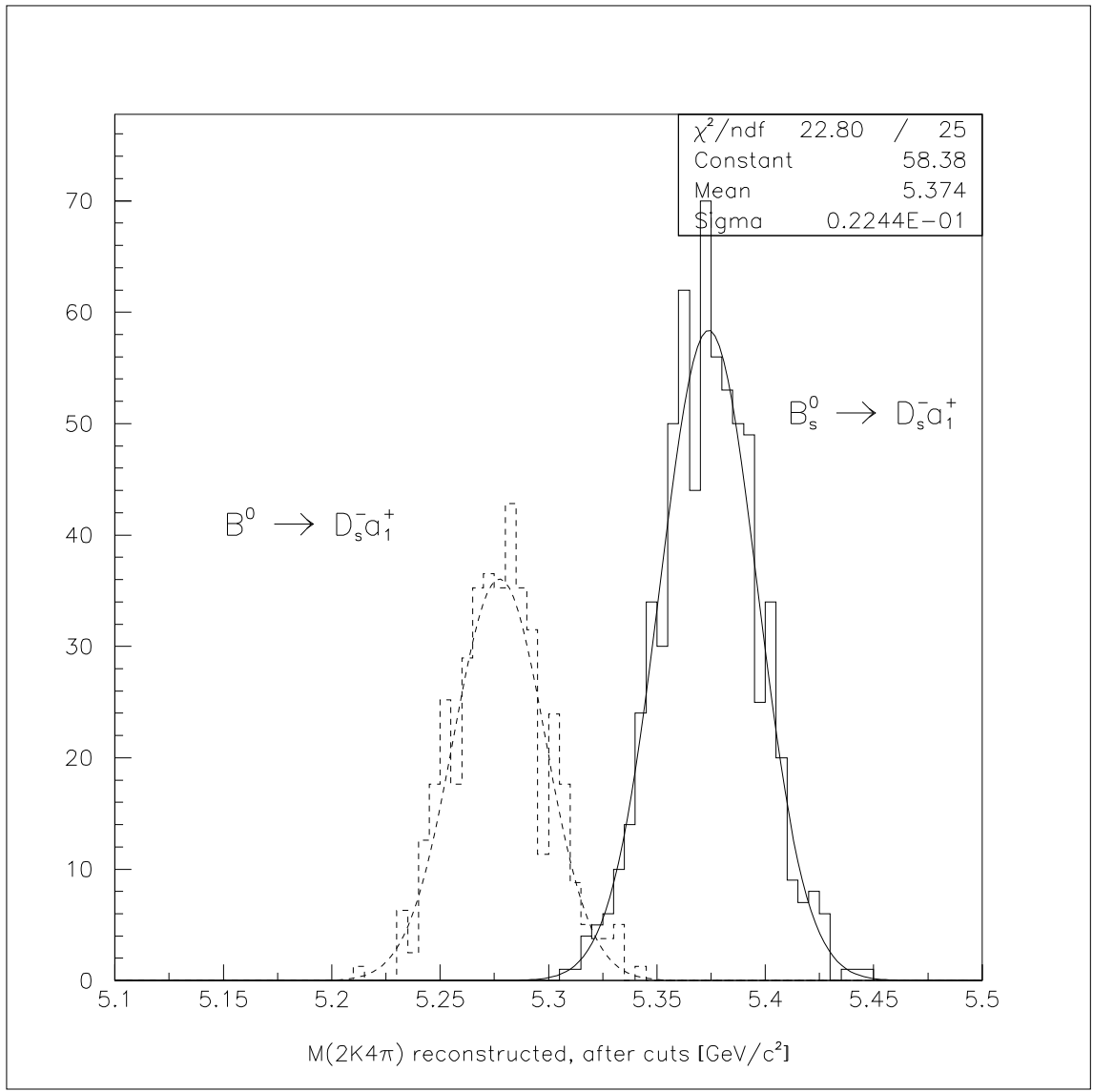,height=13cm,bbllx=40pt,bblly=70pt,bburx=625pt,bbury=550pt}
\caption{Six particle invariant mass distribution corresponding to the
$B_s^0$ meson. Dashed line - expected upper limit for background from
$B^0$ decay.}
\end{figure}

\begin{figure}
\epsfig
{file=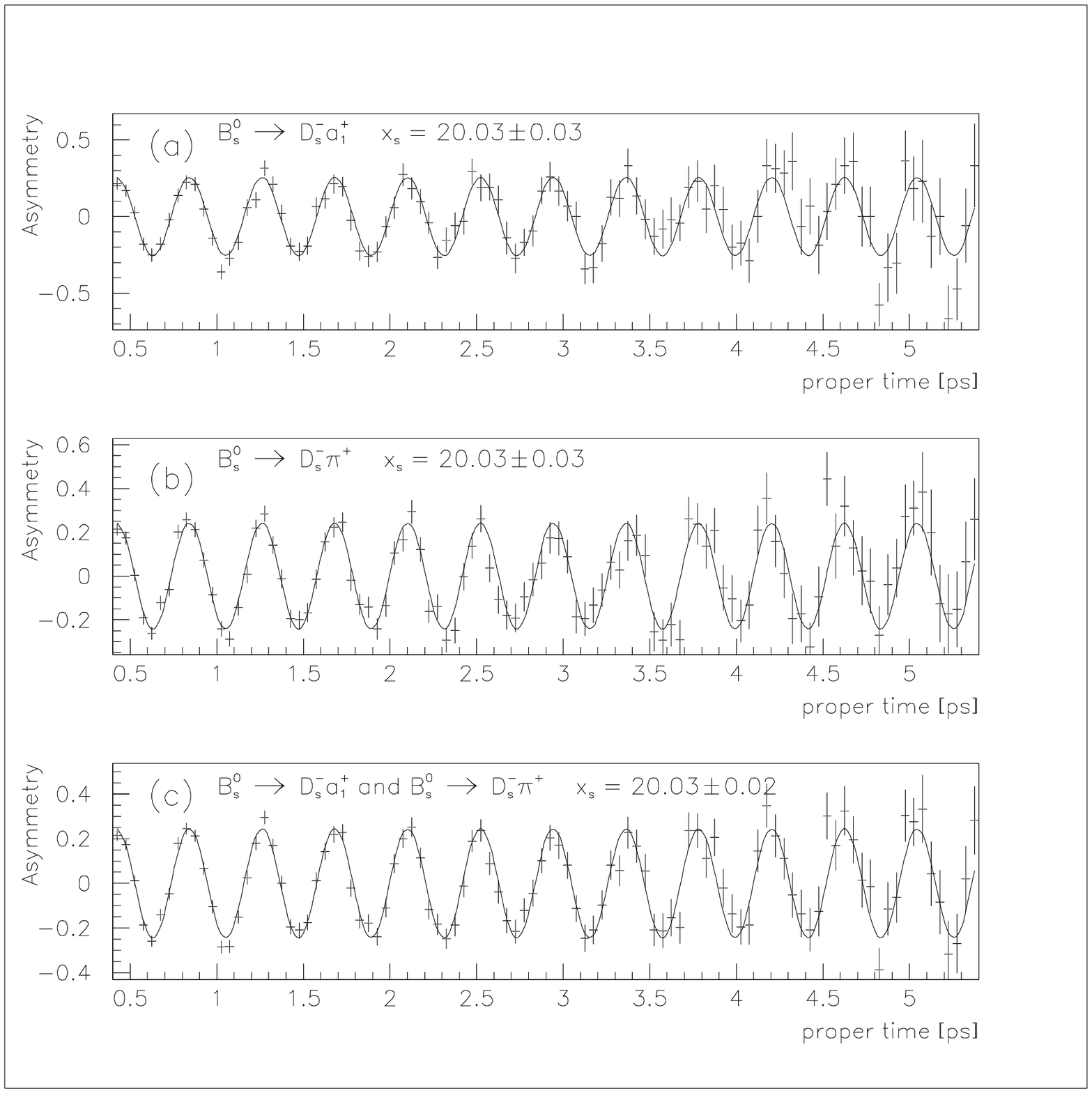,height=13cm,bbllx=40pt,bblly=70pt,bburx=625pt,bbury=550pt}
\caption{Asymmetry distributions for $B_s^0 \to D_s^- a_1^+$ (a),
$B_s^0 \to D_s^- \pi^+$ (b) and when both channels are used (c),
 for $x_s=20$.}
\end{figure}

\begin{figure}
\epsfig
{file=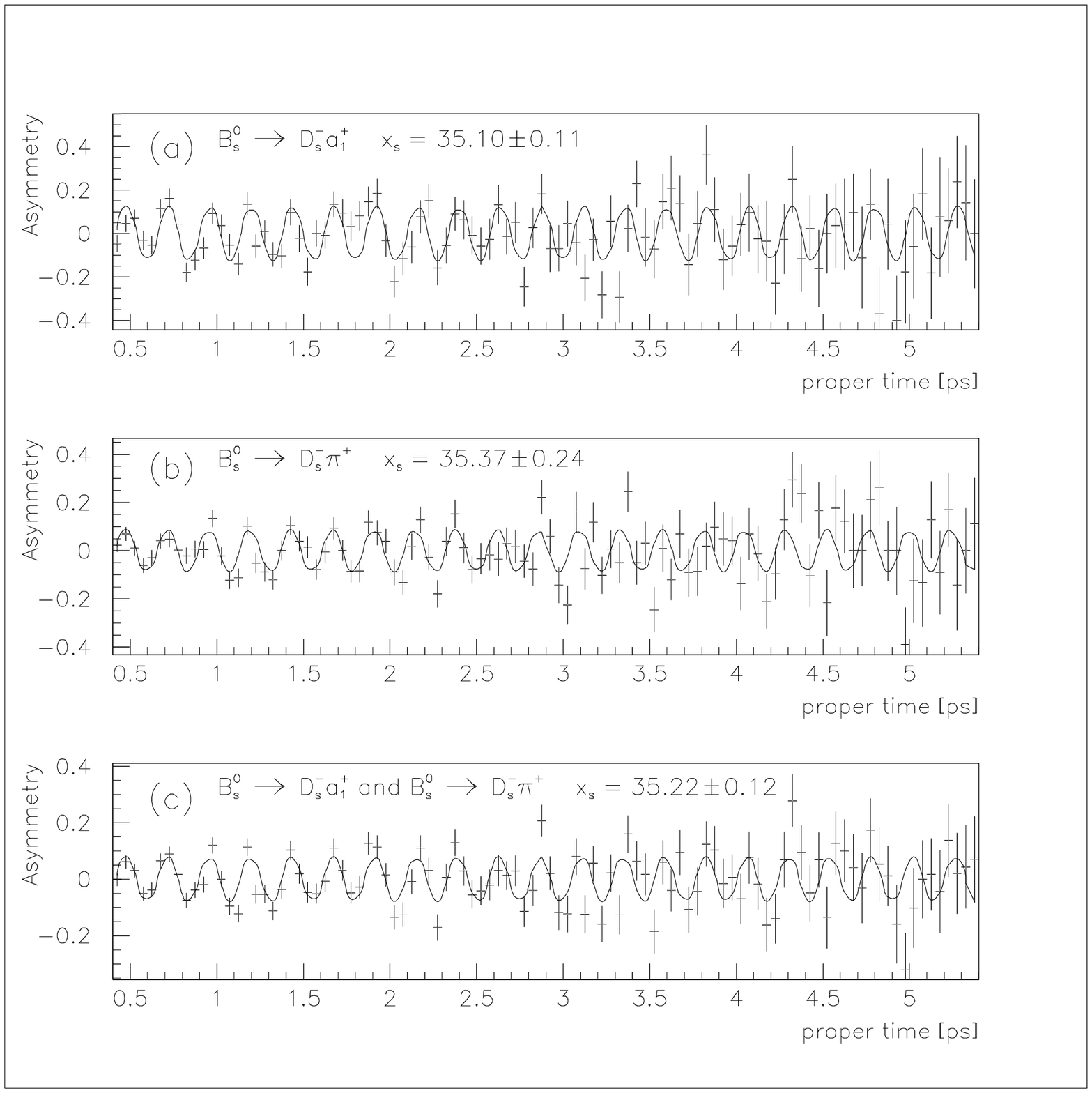,height=13cm,bbllx=40pt,bblly=70pt,bburx=625pt,bbury=550pt}
\caption{Asymmetry distributions for $B_s^0 \to D_s^- a_1^+$ (a),
$B_s^0 \to D_s^- \pi^+$ (b) and when both channels are used (c),
 for $x_s=35$.}
\end{figure}


\end{document}